\documentclass[a4paper,12pt]{article}
\usepackage{amsmath}
\usepackage[latin1]{inputenc}
\usepackage{amssymb}
\usepackage{graphicx}
\usepackage{makeidx}
\usepackage{mathrsfs}
\usepackage{amsfonts}
\usepackage[mathscr]{eucal}
\usepackage{amsmath}
\font\fmtitle=cmbx12 scaled \magstep2
\font\text=cmr10 at 12 truept
 at 11 truept
 at 10 truept

\def\pni{\par \noindent}
\def\vsh{\vskip 0.25truecm\noindent}
\def\vs{\vskip 0.5truecm}

\def\vsp{\vsh\pni}
\def\vsn{\pni}

\def\eg{{e.g.}\ } \def\ie{{i.e.}\ }

\def\e{\hbox{e}}
\def\exp{\hbox{exp}}
\def\ds{\displaystyle}
\def\dis{\displaystyle}
\def\q{\quad}

\def\rec#1{\frac{1}{#1}}
\def\log{{\rm log}\,}
\def\erf{{\rm erf}\,}     \def\erfc{{\rm erfc}\,}

\def\RR{\vbox {\hbox to 8.9pt {I\hskip-2.1pt R\hfil}}\;}
\def\CC{{\rm C\hskip-4.8pt \vrule height 6pt width 12000sp\hskip 5pt}}
\def\erf{{\rm erf}\,}   \def\erfc{{\rm erfc}\,}
\def\exp{{\rm exp}\,} \def\e{{\rm e}}
\def\ss{{s}^{1/2}}   %% for LAPLACE TRANSFORMS
\def\N{\bar N}  %%%%%%%%%%%%%
\def\ss{{s}^{1/2}} %%%%%%%
\def\lst{{\lambda \,\stt}}
\def\Et{{E_{1/2}(\lst)}}

\def\bar{\widetilde}

\begin{document}
\setcounter{page}{1}   \thispagestyle{empty}
%%%%%%% making running heads on even and odd pages %%%%%%%
 %\markboth
 %{\rm \centerline {F. Mainardi, A. Mura and F, }}
%{\rm \centerline{The functions of the Wright type in Fractional Calculus}}
\centerline{{\fmtitle Brownian motion and anomalous diffusion}}
\vskip 0.10truecm
\centerline{{\fmtitle revisited via a fractional Langevin equation}%%
 %\vskip 0.10 truecm
% \centerline{{\fmtitle a tutorial survey}%%
\footnote{Paper published in 
{\it Modern Problems of Statistical Physics},
Vol 8, pp. 3-23 (2009):
 a journal founded to the memory of
Prof. Ascold N. Malakhov, see
{\tt http://www.mptalam.org/i.html}}
}
\vskip 0.20truecm
%% \text
\centerline{{\bf Francesco MAINARDI}$^{a}$, {\bf Antonio MURA}$^{b}$, 
and {\bf Francesco TAMPIERI}$^c$ }
% \vskip 0.10truecm
\noindent
\begin{center}
$^a$  Department of Physics, University of Bologna, and INFN,
 \\  Via Irnerio 46, I-40126 Bologna, Italy;
%% \\  Tel: +39-051-20.91098; Fax: +39-051-247244
\\ E-mail: {\tt francesco.mainardi@unibo.it} 
\vskip 0.05truecm
 $^b$  CRESME Ricerche S.p.A, 
\\ Viale Gorizia 25C, I-00199 Roma, Italy;
 \\ E-mail: {\tt anto.mura@gmail.com}
 \vskip 0.05truecm
$^c$   Institute ISAC, CNR,
\\   Via Gobetti 101, I-40129 Bologna, Italy;
\\ E-mail: {\tt f.tampieri@isac.cnr.it}
%%%%%%%%%%
\vs
 \def\date#1{\gdef\@date{#1}} \def\@date{\today}
%% {\bf Ver Final, \@date }
% {PRELIMINARY VERSION \@date}
 \end{center}
%%%%%%%%%%%%%%%%%%%%%%
\noindent
{\it Keywords}: Brownian motion, Basset force, anomalous diffusion, Langevin equation, fractional derivatives. 
%%%%%%%%%%%%%%%%% SUMMARY %%%%%%%%%
%def\indice{\leaders\hbox to 1 em {\hss.\hss}\hfill}
%\def\hb#1{\hbox to 0.9 truecm{ \hss#1}}
% \section*{Contents}
\\
{\it PACS}:
% I select the 3 PACS numbers from  02: Mathematical methods in physics and 
% 3 PACS numbers from 05: Statistical physics,thermodynamics, and nonlinear dynamical systems 
%(see also 02.50.-r Probability theory, stochastic processes, and statistics)
% Candidate numbers are:
02.30.Gp, %%  (Special functions);
02.30.Uu, %% (Integral transforms);
02.60.Jh,  %%(Integral and integrodifferential equations);
05.10.Gg, % (Stochastic analysis methods:Fokker-Planck, Langevin, etc.)
05.20.Jj, %(Statistical mechanics of classical fluids)
05.40.Jc. % ( Brownian motion)
%%%%%%%%%%%%%%
 \section*{Abstract}
In this paper we  revisit the Brownian motion on the basis of {the fractional Langevin equation}  
which turns out to be a particular case of  the generalized Langevin equation  introduced by Kubo in 1966.
The importance of our approach is to model the Brownian motion more realistically than the usual one based 
on the classical Langevin equation, in that it takes into account also the retarding effects 
due to {hydrodynamic back-flow}, i.e. the added mass and the Basset memory drag. 
% \vsp
We provide the analytical expressions of {the correlation functions} (both for the random force and the 
particle velocity)  and of { the mean squared particle displacement}. 
% \vsp
The random force  has been shown to be represented by a superposition of the usual white noise with a 
{"fractional" noise}. 
% \vsp
{The velocity correlation function} is no longer expressed by a simple exponential  
but  exhibits a slower decay, proportional to $t^{-3/2}$  for long times, which indeed is more realistic.  %
% \vsp
Finally,  {the mean squared  displacement} is shown to maintain, for sufficiently long times, 
the linear behaviour  which is typical of normal diffusion, 
with the same diffusion coefficient of the classical case.  
% \vsp
However, {the Basset history force} induces
  a retarding effect  in the establishing of the linear behaviour, 
  which in some cases could appear as a manifestation of  {anomalous diffusion} to be 
  correctly interpreted in experimental measurements.

% \newpage 

%%%%%%%
\section{Introduction}
%%%%%%%%%%
Since the pioneering computer experiments by  %% Rahman (1964)
Alder \& Wainwright (1970) \cite{Alder-Wainwright 70},
which have shown that  the velocity
autocorrelation function for a Brownian particle in a dense fluid goes
asymptotically as $t^{-3/2}$ instead of exponentially as predicted by
stochastic theory, many attempts have been made to reproduce  this result
by purely theoretical arguments.
\vsp
Usually,  hydrodynamic models  are adopted to generalize Stokes' law
for the frictional force and obtain a {\it Generalized Langevin Equation} (GLE)),
see \eg
Zwanzig \& Bixon (1970, 1975) \cite{Zwanzig-Bixon 70,Zwanzig-Bixon 75},
  Widom (1971) \cite{Widom 71},
Case (1971) \cite{Case 71},
Mazo (1971) \cite{Mazo 71},
Ailwadi \& Berne (1971) \cite{Ailwadi-Berne 71},
Nelkin (1972) \cite{Nelkin 72},
Hynes (1972) \cite{Hynes 72},
Chow \& Hermans (1972a, 1972b, 1972c)
\cite{Chow-Hermans 72a,Chow-Hermans 72b,Chow-Hermans 72c},
Hauge \& Martin-L\"of (1973) \cite{Hauge-Martin-Lof 73},
Dufty (1974) \cite{Dufty 74},
Bedeaux \& Mazur (1974) \cite{Bedeaux-Mazur 74},
Hinch (1975) \cite{Hinch 75},
Pomeau \& R\'esibois (1975) \cite{Pomeau-Resibois 75},
Warner (1979) \cite{Warner 79},
Reichl (1981) \cite{Reichl 81},
Paul \& Pusey (1981) \cite{Paul-Pusey 81}, %% quoted by CLERCX \& SCHRAM
Felderhof (1991) \cite{Felderhof 91}, %% quoted by CLERCX \& SCHRAM
Clercx \& Schram  (1992) \cite{Clercx-Schram 92}.
%%%%%%%
\vsp
 Recently, a great interest on the subject matter has
been raised because of the possible connection among
long-time correlation effects, fractional Langevin equation
and anomalous diffusion,
see \eg Muralidar et al. (1990) \cite{Muralidar-et-al 90},
Wang \& Lung (1990) \cite{Wang-Lung 90},
Wang (1992) \cite{Wang 92}, 
 Porr\`a et al. (1996) \cite{Porra-et-al PRE96},
Wang \& Tokuyama (1999)\cite{Wang-Tokuyama 99},
Kobelev \& Romanov \cite{Kobelev-Romanov 00}, Metzler \& Klafter \cite{Metzler-Klafter PRE00},
Lutz (2001) \cite{Lutz PRE01},
Bazzani et al. (2003) \cite{Bazzani-et-al 03}, 
Budini \& C\'aceres (2004) \cite{Budini-Caceres 04}, Pottier \& Mauger \cite{Pottier-Mauger 04},
Coffey et al. (2004) \cite{Coffey BOOK04},
Khan \& Reynolds (2005) \cite{Khan-Reynolds 05},
Fa (2006) \cite{Fa PRE06}, Vi\~nales \& Desp\'osito \cite{Vinales PRE06},
Fa (2007) \cite{Fa EPJ07}, Vi\~nales \& Desp\'osito \cite{Vinales PRE07},
Taloni \& Lomholt (2008) \cite{Taloni-Lomholt  PRE08},
Vi\~nales et al. \cite{Vinales PRE09}, Desp\'osito \& Vi\~nales (2009) \cite{Desposito-Vinales PRE09},
Figueiredo Camargo et al. (2009a, 2009b) \cite{Figueiredo-et-al 09a,Figueiredo-et-al 09b},
Eab \& Lim (2009 ) \cite{Eab-Lim E-Print09}.
%%%%%%%%%%%%%%%%%%%%
\vsp
We recall that anomalous diffusion is the phenomenon, usually
met in disordered or fractal media see \eg Giona \& Roman (1992) \cite{Giona-Roman 92}, according to which the
mean squared displacement (the variance) is no longer linear in time (as in the standard diffusion)
but proportional to a power $\alpha $ of time with
$ 0<\alpha <1$ (slow diffusion) or $1<\alpha <2$ (fast diffusion),
see \eg  Bouchaud \&  Georges (1990) \cite{Bouchaud-Georges 90}.
In general anomalous diffusion phenomena are related to generalized diffusion equations containing 
fractional derivatives in space and/or in time and,  on this respect,  the literature is huge.
We limit to recall some (review) papers and books that have mainly attracted our attention in the last decade
to  which the reader can refer along with the references therein:
Carpinteri \& Mainardi - Editors (1997) \cite{Carpinteri-Mainardi CISM97},
Saichev \& Zaslvasky (1997) \cite{Saichev_97},
Grigolini et al. (1999) \cite{Grigolini-et-al PRE99},
Uchaikin \& Zolotarev (1999) \cite{Uchaikin-Zolotarev BOOK99},
Metzler \& Klafter (2000) \cite{Metzler-Klafter PhysRep00},
Hilfer - Editor (2000) \cite{Hilfer BOOK00}, 
Mainardi et al. (2001) \cite{Mainardi LUMAPA01}   
Gorenflo et al. (2002) \cite{Gorenflo-et-al CP02},
Saichev \& Utkin (2002) \cite{Saichev-Utkin MSPS02},
Meerschaert et al. \cite{M3-et-al PRE02}, 
Metzler \& Klafter (2004) \cite{Metzler-Klafter JPhysics04},
Piryatinska et al. (2005) \cite{Saichev PhysicaA05},
Dubkov et al. (2008) \cite{Dubkov-et-al_IJBC08}, Uchaikin  (2008) \cite{Uchaikin_BOOK08},
Klages et al. - Editors (2008) \cite{Klages-et-al BOOK08}.
\vsp
We also point out that
Kubo (1966)  \cite{Kubo 66}
introduced a  generalized Langevin equation (GLE),
where the friction force appears retarded or frequency dependent
through an indefinite memory function. %% as an integral kernel.
%% In other words, this theorem may be represented
%% by a stochastic equation describing the fluctuation, which is a
%% generalization of the classical Langevin equation;
To be consistent with the
{\it fluctuation-dissipation  theorem},
in GLE the random force is no longer a white noise (as in the classical
Langevin equation) with a frequency spectrum related to that
of the velocity fluctuations. 
%% See later: %%% A critical analysis of Kubo's derivation of the fluctuation-dissipation theorem was given
%% by Felderhof (1978) \cite{Felderhof 78}.
As a matter of fact, the hydrodynamic models appear as particular cases of GLE, as noted by
Kubo  et al. (1991) \cite{Kubo-et-al 91}.
\vsp
In this paper we  revisit the Brownian motion
via a generalized {Langevin equation} of fractional order based 
on the approach started in earlier 1990's 
by Mainardi and collaborators, see
Mainardi et al. (1995) \cite{Mainardi-et-al 95}
and Mainardi \& Pironi (1996) \cite{Mainardi-Pironi 96}.
%% The purpose of our approach is  to
We model the  Brownian motion
by taking into account not only the Stokes viscous
drag (as in the classical model) but  also the retarding effects due to
hydrodynamic back-flow, \ie the added mass, and
the Basset  history force.
%% more realistically  than the usual one based on the classical Langevin
%% equation, in that it takes into account not only the Stokes viscous
%% drag but  also the retarding effects due to
%% hydrodynamic back-flow, \ie the added mass and
%% the Basset-Boussinesq  history force.
 \vsp %%%%%%%%%
The plan of the paper is the following.
After having reviewed in Section 2 the classical Brownian motion,
in Section 3 we extend the theory according to the hydrodynamic approach.
On the basis of the  {\it fluctuation-dissipation theorem}
(see Appendix A)
and  of the techniques of the {\it fractional calculus}
 (see Appendix B),
we provide the analytical expressions
of the autocorrelation functions (both for the random force and  the
particle velocity)  and of the displacement variance.
Consequently,    %%  the  well-known results of
the classical theory
of the Brownian motion  is properly generalized.
Significant results are shown and discussed in Section 4,
where we  point out different diffusion regimes.
Finally, conclusions are drawn in Section 5.
%%%%%%
%%%%%%%
\section{The classical approach to the Brownian motion}
%%%%%%%%%%
We assume that the Brownian particle of mass $m$ executes
a random motion in one dimension with velocity      $V= V(t)$   and
displacement $X=X(t)$.
%%\vsp
The classical approach to the Brownian motion is based
on  the following stochastic  differential equation
({\it Langevin equation}), see 
\eg Kubo et al. (1991) \cite{Kubo-et-al 91},
  $$  m\, \frac{dV}{dt} =  F_v(t)  + R(t)\,,\eqno (2.1a)$$
where $F_v(t)$ denotes  the {\it frictional force} exerted from
the fluid on the particle
and $R(t)$ denotes the {\it random force} arising from rapid
thermal fluctuations.
%%%%%%%%%%%%%%%%
$R(t)$ is defined such that for any $t_1\neq t_2$ the random variables
$R(t_1)$ and $R(t_2)$ are independent and $\{R(t), t\ge 0\}$ 
defines a stationary zero-mean process. 
As usual, we denote with brackets the average taken over an ensemble in thermal equilibrium, 
i.e. $\langle \, R(t)\, \rangle= 0$.
%%%%%%%%
\vsp
Assuming for the frictional force the Stokes  expression
for a drag of spherical particle of radius $a\,,$  we
obtain the classical formula
$$ F_v = -\frac{1}{\mu } \, V(t) \,, \q
    \frac{1}{\mu }= 6 \pi \, a \,\rho_f \,\nu\, \,, \eqno(2.2)$$
where $\mu$ denotes the {\it mobility coefficient} and
$\rho _f$ and $\nu $ are the density and the kinematic viscosity
of the fluid, respectively.
If we introduce the friction characteristic time
$$ \sigma   =   m\, \mu \,,  \eqno(2.3)$$
the Langevin equation (2.1) explicitly reads
   $$  \frac{dV}{dt} =
 -\frac{1}{\sigma }\,  V(t) +  \frac{1}{m} \, R(t)\,,\eqno (2.4)$$
 which corresponds to the stochastic differential equation
$$  
dV(t) = -\frac{1}{\sigma}\,  V(t)dt + dB(t)\,,\eqno(2.4a)$$
where $B(t)$ is a suitable chosen Wiener process. Eq. (2.4a) can be easily solved and gives  
$$
V(t)=\e^{-t/\sigma}\left(Z+\int_{t_0}^t \e^{s/\sigma}dB_s \right),\;\; 0<t_0\le t\,,\eqno(2.4b)
$$
with $Z=e^{t_0/\sigma}V(t_0)$. 
From the equation above it is easy to evaluate autocovariances; 
in fact, provided that $\langle Z\rangle=0$, one has, for any $\tau, s\ge t_0$:
$$
C_V(\tau,s)=\langle V(\tau)V(s)\rangle=
\e^{-(\tau+s)/\sigma}\left\{\langle Z^2\rangle -\frac{\sigma \gamma^2}{2}
\left (\e^{-2t_0/\sigma}-e^{2\min(\tau,s)/\sigma}\right)\right\}, \eqno(2.4c)
$$ 
where $\gamma$ is such that $\langle B(t)B(s)\rangle=\gamma^2 \min(t,s)$, for any $t,s>0$.
 \vsp
 At this point one usually assumes that the Brownian particle has been 
kept for a sufficiently {\it long time} in the fluid at (absolute) temperature $T\,, $ 
so that the thermal equilibrium is reached. 
Thus,  for any $\tau,s$ in  which the thermal equilibrium is maintained, $C_V(\tau,s)$ reduces to:
$$
C_V((\tau,s))=\frac{\sigma\gamma^2}{2}e^{-\displaystyle{\mid \tau-s\mid/\sigma}}. \eqno(2.4d)
$$
The equipartition law  for the energy distribution requires that
$$ m \, \langle \, V^2(t)\, \rangle  =  k \, T, \eqno(2.5)$$
where $k$ is the Boltzmann constant, therefore $\gamma^2=2KT/\sigma m$.
\vsp
Eq. (2.4d) implies that for any $t$ such that the equilibrium is reached 
the {\it autocovariance functions} $C_V$ and $C_R$ of  the stochastic processes $V(t)$ and $R(t)\,$ 
can indeed be written as:
 $$ C_V(t_0,t) = \langle \,V(t_0)\, V(t_0 +t)\, \rangle = C_V(t)\,,
   \q t\ge 0\,,
   \eqno(2.6)$$
$$ C_R(t_0,t)=   \langle \,R(t_0)\, R(t_0 +t)\, \rangle = C_R(t)\,,
 \q t\ge 0\,,
  \eqno(2.7)$$
i.e  not depending on $t_0\,. $
Moreover, by definition, the random force is uncorrelated to  the  particle velocity, namely
$$   C_{VR}(t_0,t)=  \langle \, V(t_0) \, R(t_0+t) \, \rangle = 0\,,
\q t\ge 0\,.
  \eqno(2.8)$$
Hereafter  we shall assume $t_0 =0$. 
Because of Eqs. (2.4) and the previous assumptions, one has that for any $t\ge 0$
%% the following relevant results,
$$
C_V(t) =  \langle \, V^2(0)\, \rangle  \, {\e}^{\dis -t/\sigma }
=  \frac{kT}{m}   \, {\e}^{\dis -t/\sigma }. \eqno(2.9)$$
Moreover, up to a multiplicative constant
$$ 
C_R(t) = \frac{m^2 }{\sigma} \, {\langle \, V^2(0)\, \rangle} \, \delta (t) =
          \frac{m}{\sigma }\, k\,T \, \delta (t)\,,\eqno(2.10)$$
where $\delta (t)$ denotes the Dirac distribution.
%% \vsp 
Starting from Eqs. (2.6)-(2.7), one can define the {\it power spectra} or 
{\it power spectral densities} $I_V(\omega)$ and $I_R(\omega )\,, $ $\omega \in \RR\,, $ 
by taking the Fourier transforms of the respective autocovariance functions; i.e.
$$ I_V(\omega ) = \widehat C_V(\omega) =
 \int_{-\infty}^{+\infty}\!\! C_V(t)\, {\rm e}^{\dis -i\omega\,t}\,dt  \,, \q
   C_V(t) = \rec{2\pi}\,\int_{-\infty}^{+\infty}  \!\! I_V(\omega)\,
   {\rm e}^{\dis +i\omega\,t}\,d\omega \,, \eqno(2.11)$$
%% and
$$ I_R(\omega ) = \widehat  C_R(\omega) =
 \int_{-\infty}^{+\infty}\!\! C_R(t)\, {\rm e}^{\dis -i\omega\,t}\,dt  \,, \q
   C_R(t) = \rec{2\pi}\,\int_{-\infty}^{+\infty}  \!\! I_R(\omega)\,
   {\rm e}^{\dis +i\omega\,t}\,d\omega \,. \eqno(2.12)$$
The result (2.9)  shows that {\it the velocity  autocorrelation function  decays exponentially 
in time} with the decay constant $\sigma  \,,$ while (2.12) 
means that the power spectrum of $R(t)$ is  to be {\it white},
\ie independent on frequency, resulting 
$$ I_R(\omega ) \equiv I_R =
     \frac{m}{\sigma }\,  k\, T \,.   \eqno(2.13)$$
%%%%%%\vsp
It can be readily shown that the mean squared displacement of a particle
starting at the origin at $t_0 =0\,$  ({\it displacement variance})
is given by
$$ \langle\,X^2(t)\, \rangle=2\,  \int_0^t \!\! (t-\tau)\,
C_V(\tau)\,d\tau = 2\, \int_0^t d\tau_1 \int_0^{\tau _1}
  \!\! C_V(\tau )\, d\tau \,, \q t\ge 0\,.  \eqno(2.14)$$
For this it is sufficient  to recall that
 $X(t)=\int_{0}^t V(t')\,dt'\,,$ and  use  (2.4d).
we obtain
$$\langle\,X^{2}(t)\,\rangle= 2\, D
 \,   \left[\,t- \sigma  \left(1-{\rm e}^{\dis - t/\sigma }\right )
  \,\right]\,, \q t \ge 0\,,
\eqno(2.15)$$
 where
$$ D= \sigma\,{\langle\, V^2(0)\,\rangle} =
    \int _0^{\infty} \!\! C_V(t) \, dt
 \,. \eqno(2.16) $$
We  note from (2.15) that
$$\langle\,X^{2}(t)\,\rangle
    =  2 D\, t\,
 \left[1- (t/ \sigma)^{-1}+ EST \right]\,,
\q {\rm  as}\q   t\to \infty \,, \eqno(2.17)$$
($EST$ = exponentially small terms) so that
$$ D= \lim_{t \to \infty} \,
          \frac{\,\langle\,X^2(t)\, \rangle}{2t} \,.
   \eqno(2.18)$$
%%%%
Furthermore,   using (2.3),  (2.5) and (2.16), we recognize that
$$ D =  \frac{\sigma}{m}\, k\, T = \mu \, k\, T
   \,. \eqno(2.19)$$
The constant $D$ is known as the {\it diffusion coefficient}
and the equation (2.19) as the {\it Einstein relation}.
%%%%%%%%%%%%
\section{The hydrodynamic approach to the Brownian Motion}
%%%%%%%%%%%%
On the basis of hydrodynamics, the Langevin equation  (2.4) is
not all correct, since it ignores the effects of the added mass and
retarded viscous force, which are due to the acceleration of the
particle, as pointed out by several authors.
%% formerly pointed out in 1971 by Widom and Case.
%% Widom (1971),  Case (1971),  Landau \& Lifshitz (1987),
%% {Odar (1964),  Maxey \& Riley (1983)
 \vsp
The added mass effect
requires to substitute the mass of the particle
with the so-called effective mass,
    $ m_e =  %%%%% m + \rec{2}\, m_f =
  m\, [ 1 + \rho _f/(2\rho _p)]\,,$
where $\rho _p$ denotes the density of the particle,
see \eg Batchelor (1967) \cite{Batchelor 67}.
%% As a consequence, in order to keep unmodified the mobility coefficient
%% in the Stokes drag, we have to introduce $\sigma_e  $ such that
%% $$ \mu =  \frac{\sigma }{ m} = \frac{ \sigma_e}{ m_e} \,
%%     \Longleftrightarrow \,
%%   \sigma_e  =  \sigma \, \l(1 +\rec{2\chi  }\r) \,,
%%   \q {\rm with} \q \chi   =  \frac{\rho _p}{ \rho _f}\,.
Keeping
unmodified   the Stokes drag law, the relaxation time changes from
$\sigma = m \, \mu $ to $\sigma _e = m_e\,\mu \,$: thus
$$
 \sigma_e  =  \sigma \, \left(1 +\frac{1}{2\chi}\right) \,,
  \q {\rm with} \q \chi=\frac{\rho_p}{\rho _f}\,.
 \eqno(3.1)
$$
The corresponding Langevin equation has the form as (2.4),
by replacing
$m$ with $m_e$ and $\sigma $ with $\sigma _e\,. $
With respect to the classical analysis, it turns out that the  added mass
effect, if it were present alone,  would be  only
to lengthen the time scale ($ \sigma_e   >\sigma \,$)
in the exponentials entering the basic formulas (2.11) and (2.15)
and to decrease the  velocity  variance
$\langle V^2(0)\rangle\,, $
consistently
with  the energy equipartition law (2.5)  at the same temperature.
 Consequently, using (3.1), the diffusion coefficient  is unmodified
and turns out to be
$$   D =  \sigma _e  \, \langle V^2(0)\rangle  = \mu \, k\, T\,,
  \eqno(3.2)$$
so the Einstein relation (2.19) still holds.
\vsp
The retarded viscous force  effect is due to an additional term to the
Stokes drag, which is related to the history of the particle
acceleration.
This additional drag force,
proposed  by  Basset and Boussinesq in earlier
times, and nowadays referred to as the {\it Basset history force},
see \eg Maxey \& Riley (1983) \cite{Maxey-Riley 83}, 
reads (in our notation)
$$
F^{B}_{v}=-\frac{1}{\mu}\sqrt{\frac{\tau_{0}}{\pi}}
\int_{t_{*}}^{t}\frac{dV(\tau)/d\tau}{\sqrt{t-\tau}}d\tau,\,\,\,\,\,\, \tau_{0}
=\frac{a^{2}}{\nu},\,\,\,\,\,\, t>t_{*}\geq-\infty.
\eqno(3.3)$$
\vsp
%%%%%%%%%%%
We suggest to interpret the Basset force in the framework
of the  fractional calculus.
 In this respect, taking $t_* = 0\,, $  we write
$$
  \frac{1}{\sqrt{\pi}} \, \int_0^t
\frac{d V(\tau )/d\tau}{\sqrt{t-\tau}}\,d\tau
  =  \frac{d^{1/2}}{dt^{1/2}} \, V(t)   \,, \eqno(3.4)
$$
where  ${d^{1/2}/ dt^{1/2}}$ denotes the fractional derivative of
order $1/2$ (in the Caputo sense), see
for details Caputo (1967, 1969) \cite{Caputo 67,Caputo 69}, 
Caputo \& Mainardi (1971) \cite{Caputo-Mainardi 71},
Mainardi (1996, 1997) \cite{Mainardi CSF96,Mainardi CISM97}, 
Gorenflo \& Mainardi (1997) \cite{Gorenflo-Mainardi CISM97},
 and Podlubny (1999) \cite{Podlubny 99}.
This definition of fractional derivative differs from the standard one
(in  the Riemann-Liouville sense) available in classical textbooks on
fractional calculus, see 
\eg Oldham \& Spanier (1974) \cite{Oldham-Spanier 74}
Ross (1975) \cite{Ross 75},
Samko et al. (1993) \cite{SKM 93}
and  Miller \& Ross (1993) \cite{Miller-Ross 93}. 
In fact, if $f(t)$
denotes a causal (sufficiently well-behaved) function and $0<\alpha <1$
we have
$$
{\ds \left(\frac{d^{\alpha }}{dt^{\alpha }}\right)_{RL}\,[f(t)]=
    \frac{1}{\Gamma(1-\alpha )}\, \frac{d}{dt}\,
 \int_0^t f(\tau )\, \frac{d\tau}{(t-\tau )^\alpha}} \,, \eqno(3.5)
 $$
 %%%%%%%%%%%%$%%%%%%%%%%%%%%%%%%%%%%%%%%%%%%%%%%%%%%%%%%%%%
$$ 
 {\ds \left(\frac{d^{\alpha }}{dt^{\alpha }}\right)_{C} \,\,\,\, [f(t)]=
    \frac{1}{\Gamma(1-\alpha )}\,
         \int_0^t \,\frac{d}{d\tau} f(\tau)
		 \,\frac{d\tau}{(t-\tau )^\alpha}} 
\,. \eqno(3.6)
$$
where $\Gamma$ denotes the Gamma function,
and the suffices $RL$ and $C$ refer to Riemann-Liouville
and to Caputo, respectively.
\vsp
We`recall the relationship between the two fractional derivatives
(for a sufficiently well-behaved function):
$$
\left(\frac{d^{\alpha }}{dt^{\alpha }}\right)_{C} \, \left[f(t)\right] 
\, =\,
\begin{cases}
{\ds \left(\frac{d^\alpha}{dt^\alpha}\right)_{RL}\, \left[f(t)-f(0^+)\right]}\,, \\ \\
{\ds \left(\frac{d^{\alpha }}{dt^{\alpha }}\right)_{RL}\, \left[f(t)\right] - 
\frac{f(0^+)\, t^{-\alpha}} {\Gamma(1-\alpha)}}\,.
\end{cases}
  \eqno(3.7)$$
Then, using (2.4), (3.1) and (3.3-4), the  Langevin equation
turns out to be
$$                \frac{dV}{dt} =
 -\frac{1}{\sigma_e }\,\left[
  1 + \sqrt{\tau_0 } \,\frac{d^{1/2 }}{dt^{1/2 }}\right] \, V(t)
+  \frac{1}{m_e} \, R(t)\,,\eqno (3.8)$$
where the suffix $C$ is understood in the fractional derivative.
We agree to refer to (3.8) as to the {\it fractional Langevin equation}.
\vsp
It is worth noting that if the process is meant to be in
thermodynamic equilibrium (at $t_0=0$), we should account for the
hydrodynamic interaction long memory, and thus it is
correct to integrate the Langevin equation (3.8) from $t_* = -\infty\,. $
Basing on the observations by Dufty (1974) \cite{Dufty 74}
 and Felderhof (1978) \cite{Felderhof 78},
we introduce in (3.8) the random force
%%%%%%%%%%%%%%%%%%
%% To do that, the random force in (3.8) may be written
%% according to Dufty (1974), see also Felderhof (1978)
%%%%%%%%%%%%%%%%%%%%%%%%%%%%
$$ R_*(t) =  R(t) -     \frac{1}{\mu}\,
          \sqrt{\frac{\tau _0}{\pi}}  \,   \int_{-\infty}^0
\frac{d V(\tau )/d\tau }{\sqrt{t-\tau}}\,d\tau \,.      \eqno(3.9)
$$
\vsp
In view of the {\it fluctuation-dissipation theorem},
Kubo (1966) \cite{Kubo 66}
considered a generalized
Langevin equation (GLE), introducing a memory function
$\gamma (t)$      to represent a generic retarded effect
for the friction force.
%% For the {\it causal} case,
In our case
 Kubo's GLE reads  (in our notation)
 $$   \frac{dV}{dt} = -\int_0^t
  \gamma (t-\tau) \, V(\tau ) \, d\tau   + \frac{1}{m_e}\, R_*(t)\,,
\eqno (3.10)$$
where $R_*(t) = R(t) - m_e\, \int_{-\infty}^0
  \gamma (t-\tau) \, V(\tau ) \, d\tau \,. $
%%%%%%%%
\vsp
For the sake of convenience, from now on we shall drop the suffix
$*$ in the
Langevin equations. Basing on the fundamental hypothesis (2.8). \ie
$   \langle \, V(0) \, R(t) \, \rangle = 0\,,\; t>0\,,$
and using  the Laplace transform,
$$ f(t) \div  \bar f(s) =  \int _0^\infty
   \e^{-st}\, f(t)\, dt \,, \q s\in \CC\, $$
(where the  sign $\div$ denotes the juxtaposition of
a function depending on $t$ with its Laplace transform depending on $s\,$),
the {\it fluctuation-dissipation  theorem} is readily expressed,
according to  Mainardi \& Pironi (1996) \cite{Mainardi-Pironi 96},  
as
$$ \bar C_{{V}}(s) =  \bar{{\langle\,  {V}(0) \, {V}(t)\, \rangle}}
   =  \frac{ \langle \, {V}^2(0)\, \rangle}{ s + \bar \gamma (s)}
     \,, \eqno(3.11)$$
%%  and
$$
 \bar C_{R}(s) =  \bar{{\langle\,  {R}(0) \, {R}(t)\, \rangle}}
=  m_e^2 \,   { \langle \, {V}^2(0)\, \rangle}\,
     \bar \gamma (s)\,. \eqno(3.12)$$
For the proof of (3.11)-(3.12) see Appendix A.
\vsp
 We understand that the {\it causal} approach to Kubo's
 fluctuation-dissipation theorem can be  controversial.
 For a criticism to the causal approach we refer the reader to a
 paper by Felderhof (1978) \cite{Felderhof 78}.
%% B.U. Felderhof: On the derivation of the fluctuation - dissipation theorem,
%% {\it J. Phys. A.: Math. Gen.}, Vol. 11, No 5,pp. 921-927, 1978.
%%%%%%%%%
\vsp
The classical results  are easily recovered for $t>0$ noting that,
in the absence of added mass and history effects,
we get
$ \bar\gamma(s)=1/\sigma \,\div\, \gamma(t)=\delta(t)/\sigma \,.$
%%%%%%%%%%%%%%%%
%%%%%%%%%%%%
\vsp
For  our fractional Langevin equation (3.8), we note that
$$ \bar \gamma(s) =   \frac{1}{\sigma_e}\,
     \left[ 1 + \sqrt{\tau_0 } \, s^{1/2} \right]
        \,  \div \,
\gamma (t) = \frac{1}{\sigma_e}\,
 \left[ \delta (t) - \sqrt{\tau_0 }\frac{t^{-3/2}}{2 \sqrt{\pi}}\,\Theta(t)\right]
 \,, \eqno(3.13) $$
where $\Theta(t)$ is the Heaviside step function.
Therefore the expression for $\gamma (t)$ turns out to be defined
only in the sense of  distributions. Specifically, $\delta (t)$
is the well-known Dirac delta function
and  $t^{-3/2} \, \Theta(t)$ is the linear functional over
test functions, $\,\phi(t)\,, $ such that
$$
 { \langle \, t^{-3/2}\,\Theta(t)\,,\, \phi(t) \rangle}  \,
  = \int _0^\infty \frac{[\phi(t)-\phi(0)]}{ t^{3/2}}\, dt\,. $$
For more details on distributions,  see \eg
Gel'fand \& Shilov (1964) \cite{Gelfand-Shilov 64}
 or Zemanian (1965) \cite{Zemanian 65}.
\vsp   %%%% preso dal CISM !!!!!!
The significant change  with respect to the classical case results
from the $\,t^{-3/2}\,$ term. Not only does it imply
a non-instantaneous relationship between the force and the
velocity, but also it is a slowly decreasing function so
that the force is effectively related to the velocity over
a large time interval.
The representation of the force in terms of distributions,
as required by the $GLE$, is not
strictly necessary since
we can use  the equivalent fractional form.
\vsp     %%%% FINE CISM
Let us now consider the correlation for the random force. The inversion
of the Laplace transform $\bar C_R (s)$ yields,
using Eqs. (3.12)-(3.13),
$$ C_R (t) =
 %%%% !! to be corrected \frac{m_e^2}{\sigma_e  } \,
 \frac{m_e^2}{\sigma_e  } \,
  { \langle \, V^2(0)\, \rangle}  \,
\left[ \delta (t) - \sqrt{\tau_0 } \, \frac{t^{-3/2}}{\sqrt{\pi}}
 \,\Theta(t)
\right]
 \,,  \eqno(3.14)$$
Thus, from the comparison with the classical result (2.12),
we recognize that,  in the presence of the history force,
the random force cannot
be longer represented uniquely by a white noise;  an
additional "fractional" or "coloured"   noise is  present due to the term
$t^{-3/2}$  which, as formerly noted by
Case  (1971) \cite{Case 71}, is to be interpreted in the generalized sense
of distributions.
\vsp
Let us now consider  the velocity autocorrelation.
Inserting (3.13) in (3.11), it turns out
$$ \bar C_V (s) =
    \frac{ {\langle \, V ^2(0)\, \rangle} }{
      s +\left[1+ \sqrt{\tau_0 } \, s^{1/2}\right]/ \sigma_e }
 = \frac{ {\langle \, V ^2(0)\, \rangle} }{
      s + \sqrt{\beta /\sigma_e } \, s^{1/2} +1/\sigma_e  }
\,, \eqno(3.15)$$
where, because of (2.2-3) and (3.1), (3.3),
$$ \beta =
       \frac{\tau_0}{\sigma_e  } =  \frac{9}{2\chi   +1} =
     \frac{9 \rho _f}{2 \rho _p + \rho _f}
\,. \eqno(3.16)$$
%%%%%%%%%
We note from (3.16) that $ 0 < \beta < 9$, the limiting
cases occurring  for $\chi  = \infty$  and $\chi   =0$,
 respectively.
We also recognize that the effect of the
Basset force is expected to be negligible for $\beta \to 0\,,$
\ie  for particles
which are  sufficiently
heavy with respect to the fluid ($\rho _p \gg \rho _f$).
In this case we can assume the validity of the classical result
(2.11).  %%  for $t>t_0=0\,. $
%%%%%%%%%%%%%%%%
\vsp
Applying in (3.15) the asymptotic theorem for Laplace transform as
$s \to   0\,, $  see \eg Doetsch (1974), \cite{Doetsch 74}
we get as $t \to \infty\,, $
$$ C_V (t) \sim
    {\langle \, V ^2(0)\, \rangle}\,
 \sqrt{\beta /( 4  \pi)}
 \,   \left(t/ \sigma_e \right)^{-3/2}\,, \q t\to \infty
    \,. \eqno(3.17)$$
The presence of such a long-time tail
was  observed by \cite{Alder-Wainwright 70} in a computer
simulation of velocity correlation functions.
Furthermore, from  (3.15) it is easy to obtain
the following results
$$ C_V (0) = \lim_{s \to \infty}  s\, \bar C_V (s)
  =  {\langle\,V ^2(0)\, \rangle} \,, \qquad
 \int _0^{\infty} \!\! C_V (t) \, dt = \bar C_V (0)
 = \sigma_e \, {\langle\,V ^2(0)\, \rangle}
   \,. \eqno(3.18)$$
%% \vsp
The explicit inversion of the Laplace transform in (3.15)
can be obtained basing on Appendix B,
see also Mainardi et al. (1995) \cite{Mainardi-et-al 95},  
Mainardi (1997) \cite{Mainardi CISM97},
and reads
%% DEFINITIONS of ERROR and EXPONENTIAL FUNCTIONS %%%%%
\def\erf{{\rm erf}\,}   \def\erfc{{\rm erfc}\,}
\def\exp{{\rm exp}\,} \def\e{{\rm e}}
\def\ss{{s}^{1/2}}   %% for LAPLACE TRANSFORMS
%%%%%%%%%%%%%%%%%%%%%%%%%%%%%%%%%%%%%%%%%%%%%%%%%%%%%%%
$$
   \frac{C_V (t)}{{\langle \, V^2(0)\, \rangle}}
  = \left\{
  \begin{array}{llll}
  \displaystyle{\frac{a_+E_{1/2} (a_+\,\sqrt{t}) - a_- E_{1/2} (a_-\,\sqrt{t})}{ a_+ -a_-}},
   & \displaystyle a_{\pm}=\frac{-\sqrt{ \beta } \pm (\beta -4)^{1/2}}
  {2\sqrt{\sigma_e  }},\\[0.5 cm]
  %% \q (\beta  \ne 4)
  E_{1/2} (a\,\sqrt{t}) \, \left[ 1 +2\, a^2\,t \right]+2a\sqrt{{t/ \pi}} 
  & \displaystyle a =-\frac{1}{\sqrt{\sigma_e}} \q (\beta  = 4),
  \end{array}
  \right.
  \eqno(3.19)
$$
where
    $$ E_{1/2} (a\,\sqrt{t})  =
   \sum_{n=0}^\infty  \frac{a^n\, t^{n/2}}{\Gamma (n/2+1)}
     =   {\rm e}^{\ds a^2\,t}\, {\rm erfc} (-a\,\sqrt{t}) \eqno(3.20) $$
denotes the {\it Mittag-Leffler  function} of order $1/2\, $
and erfc the complementary error function.
%%%%%%%%%%%%%%%%%%%
For properties of the Mittag-Leffler function we refer the reader
to Erd\'elyi (1955) \cite{Bateman HTF3},  
Gorenflo \& Mainardi (1997) \cite{Gorenflo-Mainardi CISM97}, 
Podlubny (1999) \cite{Podlubny 99} 
and Mainardi \& Gorenflo (2000)\cite{Mainardi-Gorenflo JCAM00}.
%% Furthermore, for $t>0\,, $
%% $C_V(t)$  turns out to be a decreasing  function,
%% completely monotonic, \ie $(-1)^{n}\,C_V^{(n)}(t)>0\,,\,n=0,1,2,\dots$
%%%%%%%%%%%
\vsp
%%%%%%%%%%%%
Let us now consider the displacement variance, which is provided by
(2.14). From the Laplace transform
  $\bar{{\langle\,X^2(s)\, \rangle}}= 2\, \bar C_V (s)/s^2\,, $
we derive the asymptotic behaviour  of
 ${\langle\,X^2(t)\, \rangle}$
as $t\to \infty\,, $  which reads
$$ {\langle\,X^2(t)\, \rangle}  =
2 D\, t \left\{1 -
 2\sqrt{\beta / \pi}\,  (t/\sigma _e) ^{-1/2} +
   O\left[ (t/\sigma _e)^{-1} \right] \right\}
\,,  \q t\to \infty \,, \eqno(3.21)$$
where  $D$   is the diffusion coefficient (3.2).
%%%%%%%
The explicit expression of the displacement variance turns out to be,
 taking $\beta \ne 4\,, $
$$
{\langle\,X^2(t)\, \rangle}  =  2 D
  \left\{ t-  2\sqrt{\frac{\beta \sigma _et}{\pi}} +
\frac{a_+^3 \,[1- E_{1/2}(a_- \sqrt{t})] -a_-^3 \,[1- E_{1/2}(a_+ \sqrt{t})]}{(a_+-a_-)\, (a_+\, a_-)^2} \right\}. 
 \eqno(3.22)$$
%% \vsp
Thus,  the   displacement  variance is proved
to maintain, for sufficiently long times,
the linear behaviour  which is typical of normal diffusion
(with the same diffusion coefficient  of the classical case).
However, the Basset history force,
which is responsible of the algebraic decay of the velocity correlation
function, induces  a retarding effect ($\propto t^{1/2}$) in the
establishing   of the linear behaviour.
%%%%%%%%%%%%%%%%%%%%%%%%%%%%%% SECTION 4 %%%%%%%%%%%%%
\section{Numerical results and discussion}
%%%%%%%
In order to get a physical insight of the effect of the
Basset history force we exhibit some plots
concerning the  velocity autocorrelation (3.19) and the
displacement variance (3.22), for some values of the characteristic
parameter $\chi=  \rho _p/\rho _f\,. $
%% As an example we consider relatively light Brownian particles,
%% by assuming  $\chi= 0.1\,$ and $\chi= 0.5\,. $
As we have noted before,  
the Basset  retarding effect is  more relevant
when the parameter $\beta$ introduced in (3.16) is big enough,
namely when $\chi := \rho _p/\rho _f$ is sufficiently small.
In this section, devoted to numerical results and discussion, the plots
will clearly show the increasing effect as $\chi$ becomes smaller and smaller. 
But in order to recognize the type of anomalous diffusion induced by  the behaviour of the 
{displacement  variance} what can we do?
We postpone  our approach to this problem to the final discussion.
% where we show a number of plots that would point out long  regimes
% of anomalous diffusion that would be interpreted  in experiments  as
% fast diffusion.    
\vsp
We now agree to take {\it non-dimensional} quantities,
by scaling  the time with the  decay constant $\sigma$
of the classical Brownian motion and the displacement
with the diffusive scale $(D\,\sigma )^{1/2}\,. $
With these scales the asymptotic equation for the displacement variance
reads ${\langle\,X ^2(t)\, \rangle} \sim 2 \,t\,. $
\vsp
For decreasing values of $\chi = 1, 0.5, 0.1 $ we consider the velocity autocorrelation
normalized with its initial value
 ${\langle\,V ^2(0)\, \rangle} \,$
and the displacement variance normalized with its asymptotic
value $2\,t\,.$
In Figure 1 we compare versus time the functions $C_V$ and
${\langle\,X ^2\, \rangle}/(2t) \,$
provided by our full
hydrodynamic approach (added mass and Basset force), in continuous
line, with  the corresponding ones,
provided by the classical analysis, in dashed line,
and by  the only effect of the added mass, in dotted line.
For large times we also exhibit the asymptotic estimations (3.17)  and
(3.21), in dotted line, in order to recognize  their range of validity.
\vsp
The correlation plots exhibit the well-known algebraic tail.
The time necessary to reach the asymptotic behaviour increases
as the density ratio $\chi$ decreases. By comparing the two figures,
it appears that the variance approaches the asymptotic regime
as the autocorrelation becomes sufficiently small,
independently on its time dependence.
% \newpage
%%%%%%%%%%%%%%%%%%%%%  INCLUDE GRAPHICS
\begin{figure}[htbp]
\begin{center}
 \includegraphics[width=.48\textwidth]{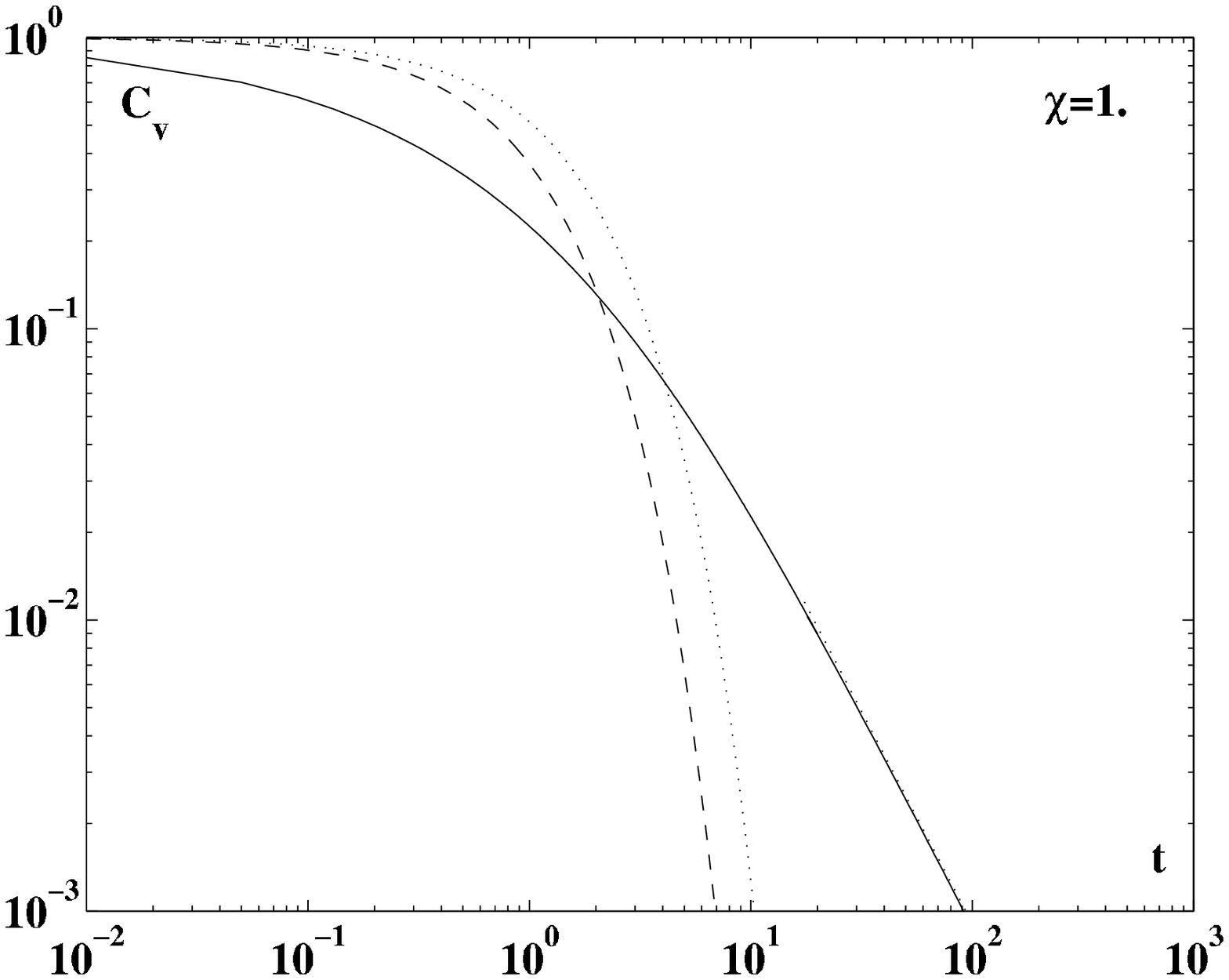}
\includegraphics[width=.48\textwidth]{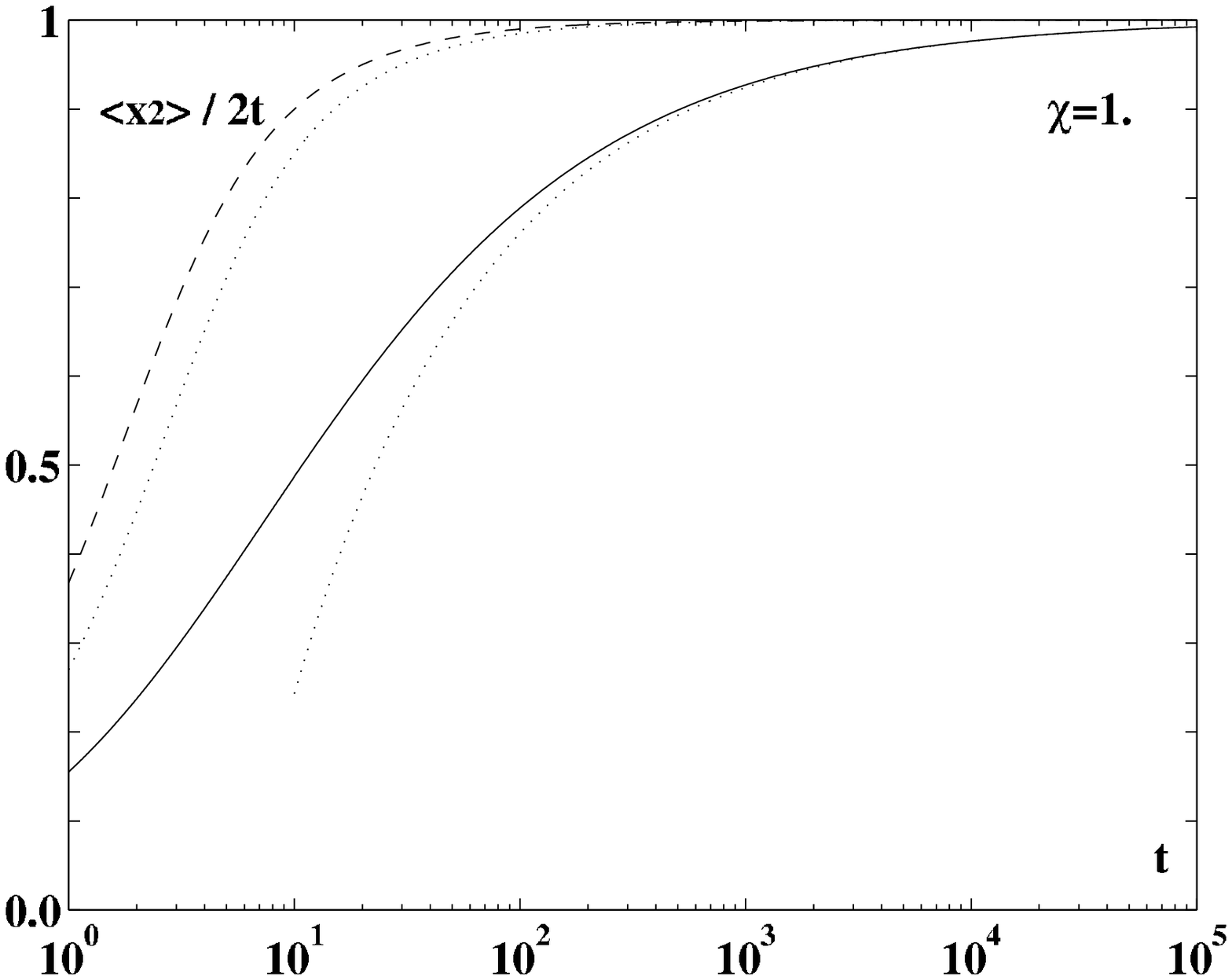}
\vskip 0.5truecm
 \includegraphics[width=.48\textwidth]{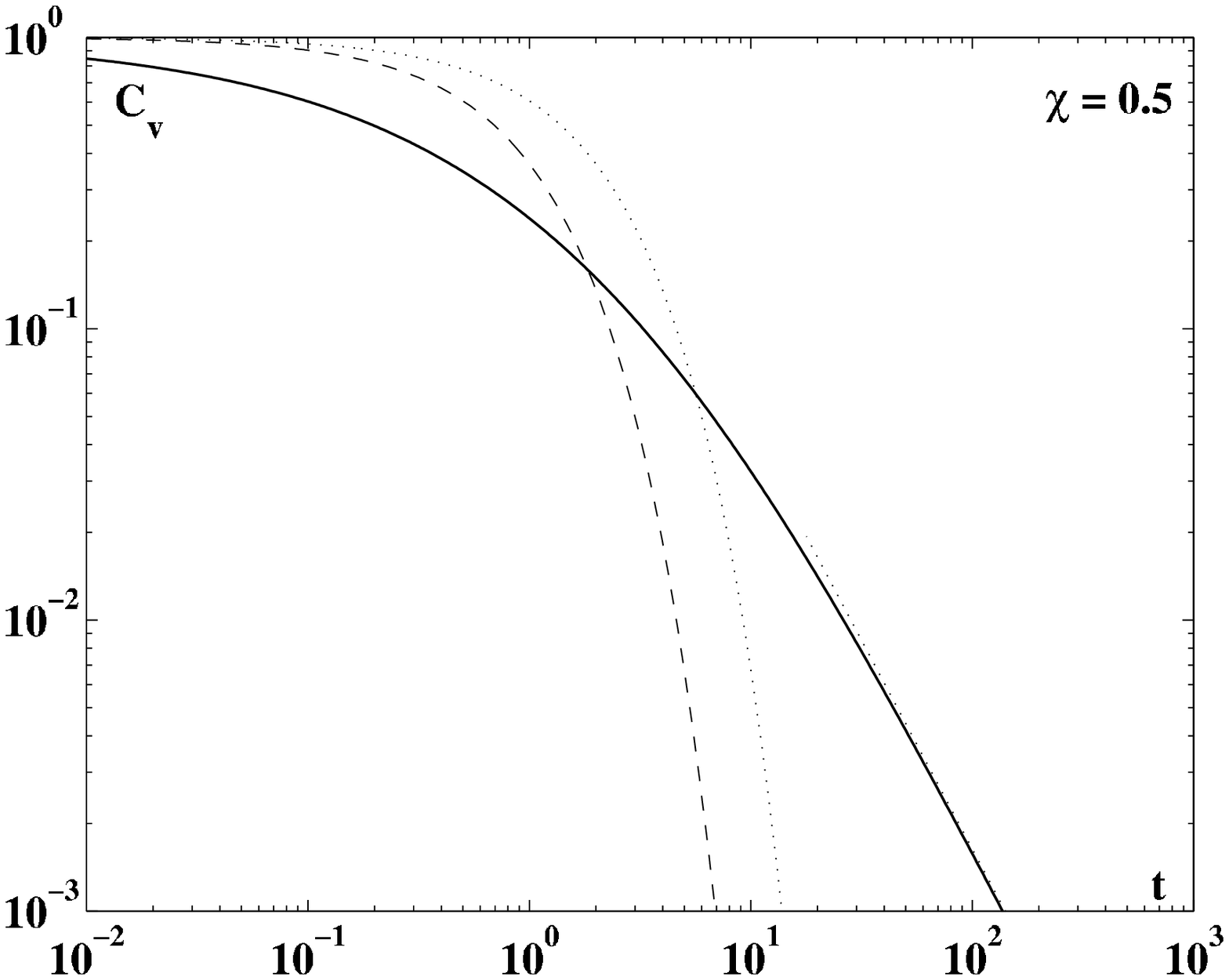}
\includegraphics[width=.48\textwidth]{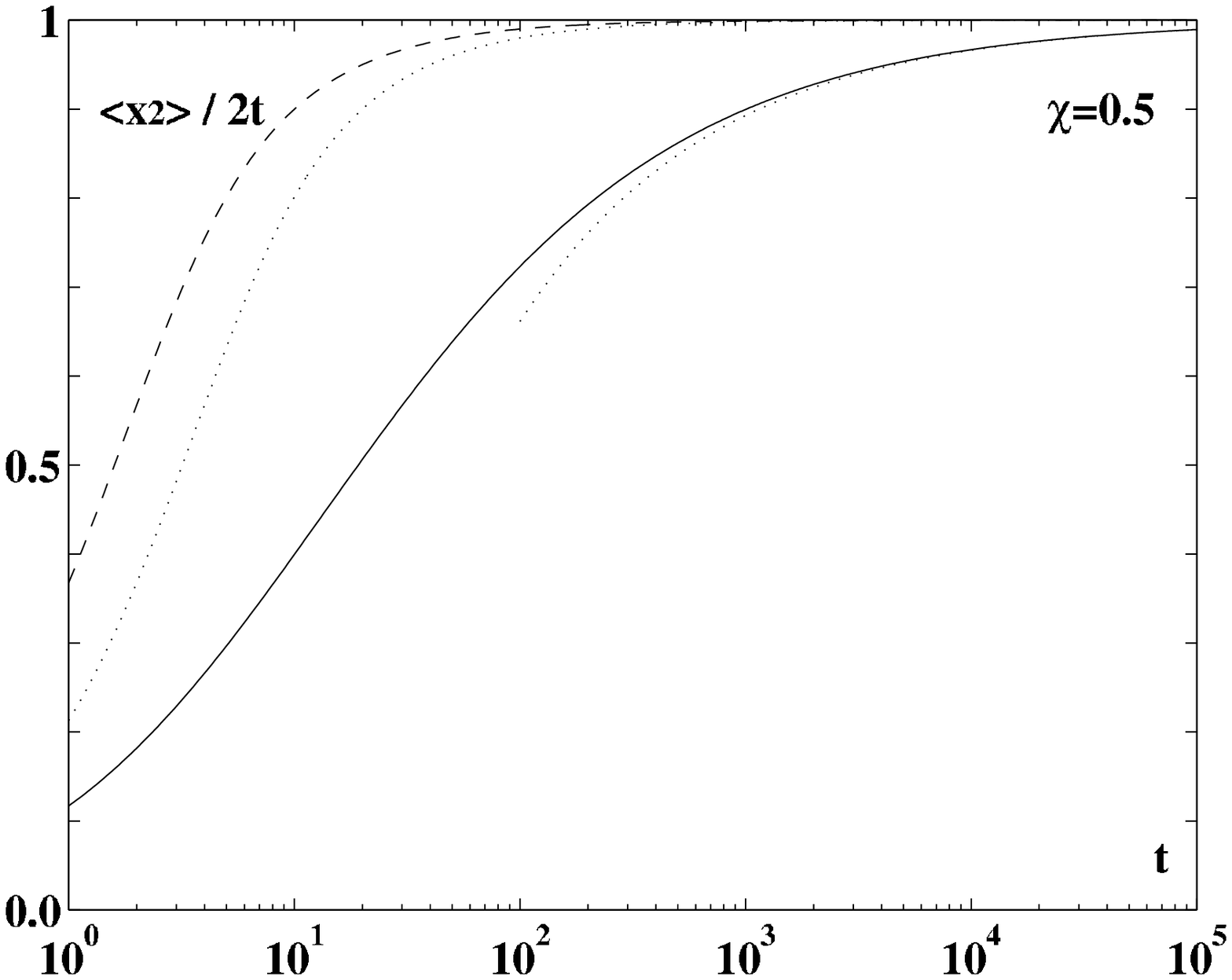}
\vskip 0.5truecm
 \includegraphics[width=.48\textwidth]{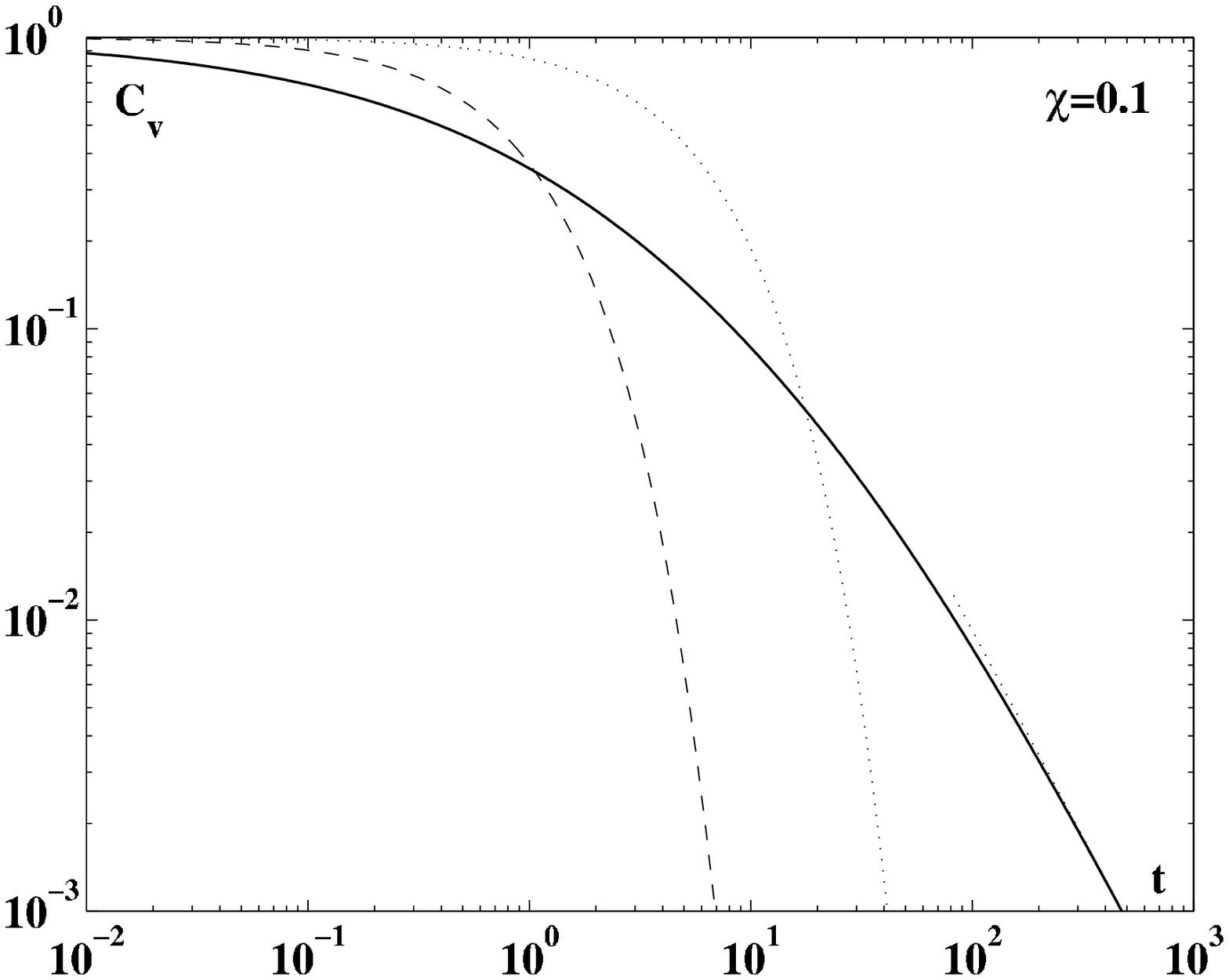}
\includegraphics[width=.48\textwidth]{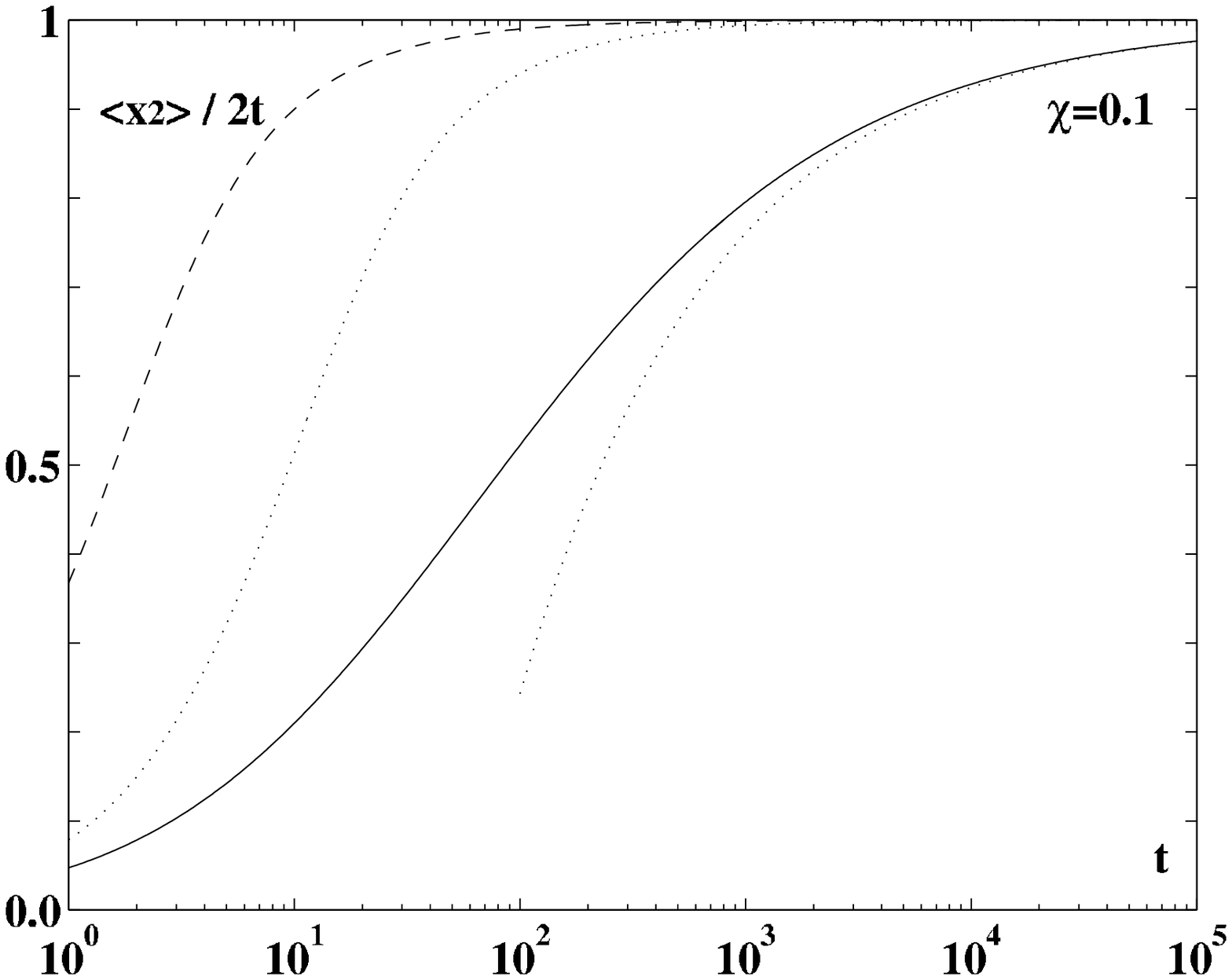}
\vskip -0.5truecm
\end{center}
 \caption{From top to bottom plots  on the velocity correlation and on the displacement variance 
 for $\chi =1, 0.5, 0.1$}
 \label{figure 1}
 \end{figure}
%\vsp
\noindent
We note that in the time interval necessary to reach the asymptotic behaviour the
displacement variance exhibits a marked deviation from the
standard diffusion.
Because this time interval turns out to be orders of magnitude longer than the classical one, 
it is relevant to discuss about various diffusion regimes before the normal one is established  
 and consequently about the nature  of the anomalous diffusion.
\vsp
In order to explore the nature of the anomalous diffusion 
we follow the following reasonings.
At first, we find it convenient to introduce  the {\it instantaneous diffusion coefficient}
%% defined as
$$ 
D(t) :=   \frac{1}{2}\, \frac{d}{dt} {\langle\,X^2(t)\, \rangle} \,.
\eqno(4.1)$$
We note that such coefficient is suitable to be measured in experiments.
 \vsp
 If we assume that ${\langle\,X^2(t)\, \rangle} $ 
varies in a range of $t$, say $t_k \le t \le t_{k+1}$, 
approximatively  as a   power law
$$
 { 
{\langle\,X^2(t)\, \rangle}  \simeq  a\, t^{\alpha_k}\,, \quad  t_k \le t \le t_{k+1}\,,
} 
\eqno(4.2)$$
where $a$ is assumed to be a suitable constant of dimensions $L^2 \, T^{-\alpha}$, called
{\it effective diffusion coefficient}, 
we are usually led to say that the diffusion regime is {\it anomalous} if $\alpha \ne 1$, precisely  
%% precisely we say
\begin{itemize}
\item {\it slow diffusion} or {\it sub-diffusion} if $\, 0<\alpha_k <1\,,$
\item {\it fast diffusion} or {\it super-diffusion} if $\, 1<\alpha_k < 2\,.$
\end{itemize}
When $\alpha_k=2$  the diffusion regime is said to be {\it ballistic}; and, of course,
when $\alpha_k=1$  the diffusion is {\it normal}.
\vsp
We easily recognize from Eq. (4.2) that the power $\alpha_k$  can be written as 
  $$ 
  { 
  \alpha_k= \frac{d (\log {\langle\,X^2(t)\, \rangle)}} {d (\log t)},\;\; t_k \le t\le t_{k+1}.
  } 
  \eqno(4.3)
  $$
  In fact from (4.2) we get for any $t_k\le t\le t_{k+1}$
 $$
\log {\langle\,X^2(t)\, \rangle}  = 
 \log \left[ a\, t^{\alpha_k} \right] = 
 \log a + \alpha_k\, \log t \Longrightarrow \; (4.3)$$
Then, varying $k$ we are able to define a function $\alpha(t)$, 
representing a sort of instantaneous diffusion parameter, 
which expresses, locally, the characterization of the regimes of anomalous diffusion.   
% \vsp
 In the following plots we shall exhibit $\langle\,X^2(t)\, \rangle$ as given by (3.22)
% $$ {\langle\,X^2(t)\, \rangle}  =  
% 2 D\left\{ t-  2\sqrt{\frac{\beta \sigma _et}{\pi}} + 
% \frac{a_+^3 \,[1- E_{1/2}(a_- \sqrt{t})] -a_-^3 \,[1- E_{1/2}(a_+ \sqrt{t})]}{(a_+-a_-)\, (a_+\, a_-)^2}\right\}.  
% \eqno(3.22)$$
 with $D(t)$ as given by (4.1) and in addition
 $\log \langle\,X^2(t)\, \rangle $ with  $\alpha(t)$ as given by (4.3).
 %%%%%%%%%
 \begin{figure}[h!]
\begin{center}
 \vskip -0.25truecm
 \includegraphics[width=.98\textwidth]{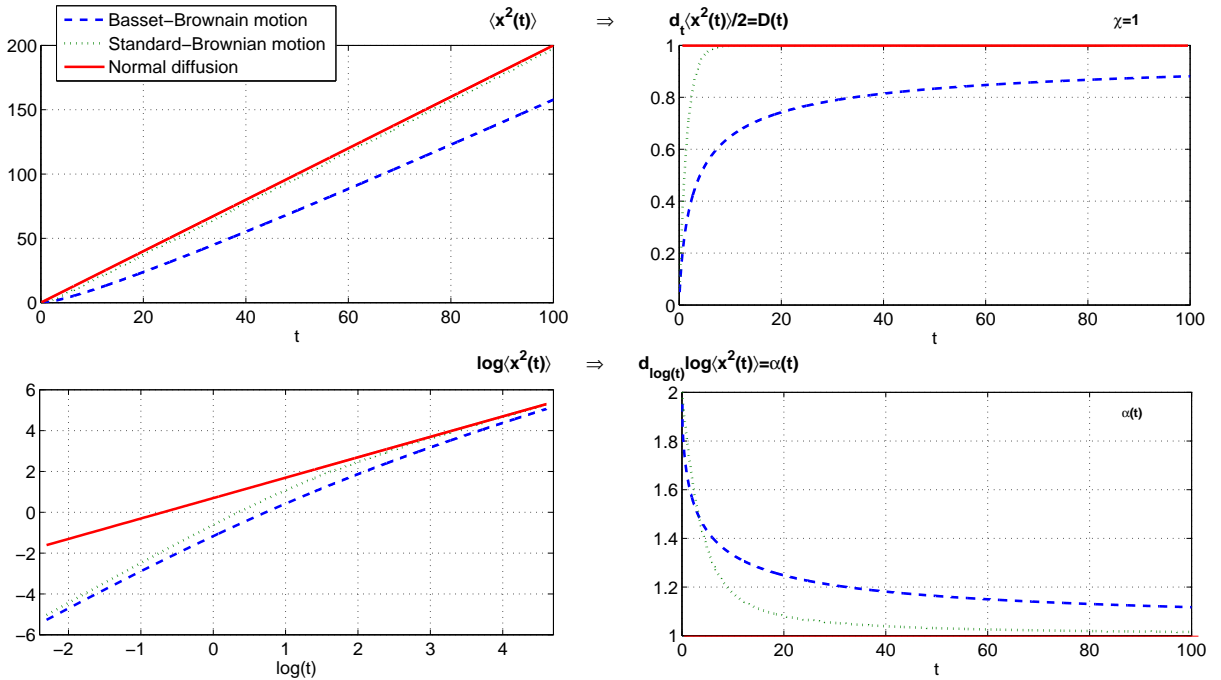}
\end{center}
\vskip -0.75truecm
 \caption{Plots on the displacement variance for $\chi =1$}
 \label{figure 2}
 \end{figure}
 %%%%%%%%%%%%%%%%%%
\newpage
%%%%%%%%%%%%%%
\begin{figure}[h!]
\begin{center}
\vskip -0.25truecm
\includegraphics[width=.98\textwidth]{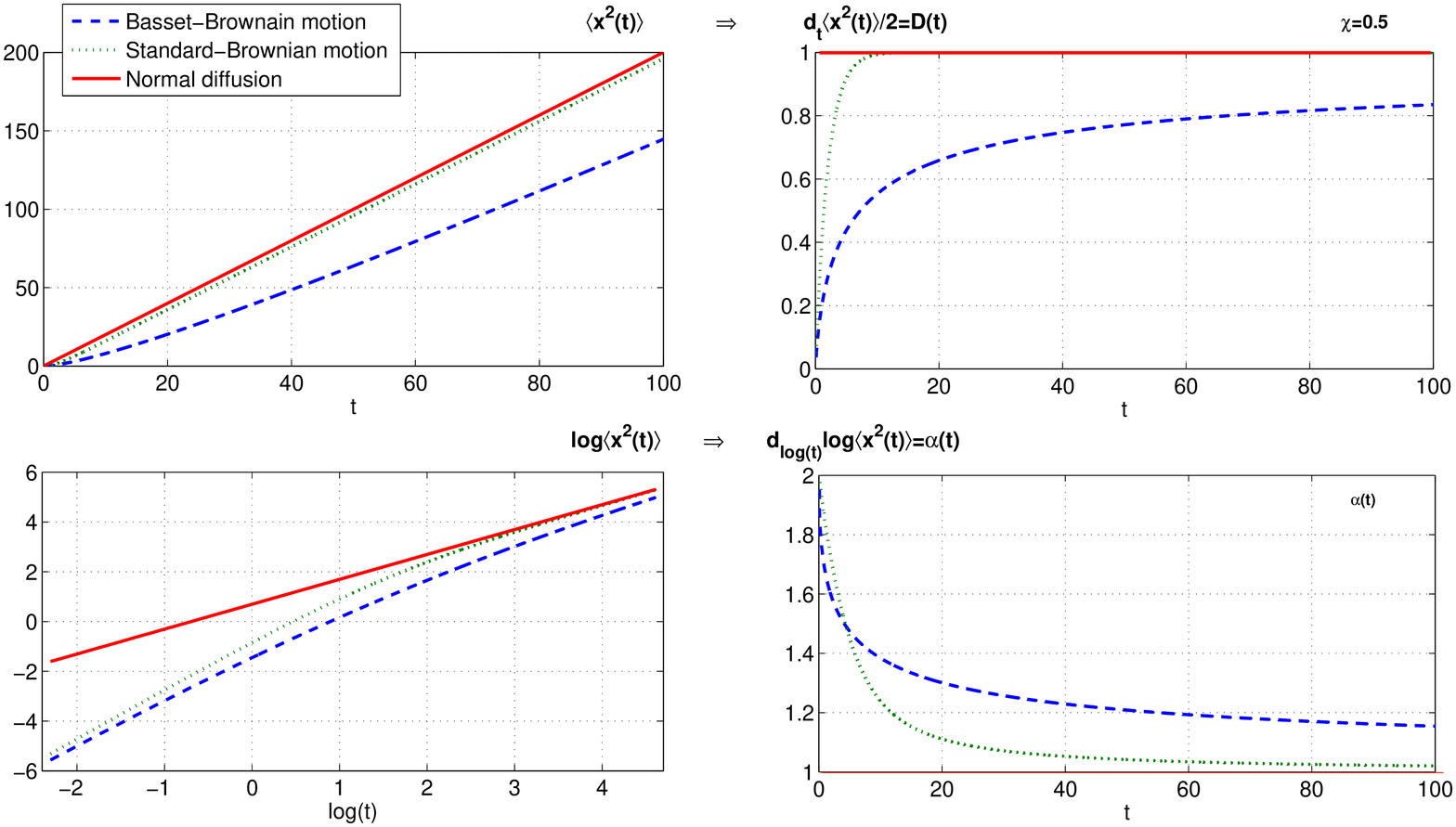}
\end{center}
\vskip -0.75truecm
 \caption{Plots on the displacement variance for $\chi =0.5$}
 \label{figure 3}
 \end{figure}
% \newpage
%%%%%%%%%%%%%%
\begin{figure}[h!]
\begin{center}
 \vskip -0.25truecm
\includegraphics[width=.98\textwidth]{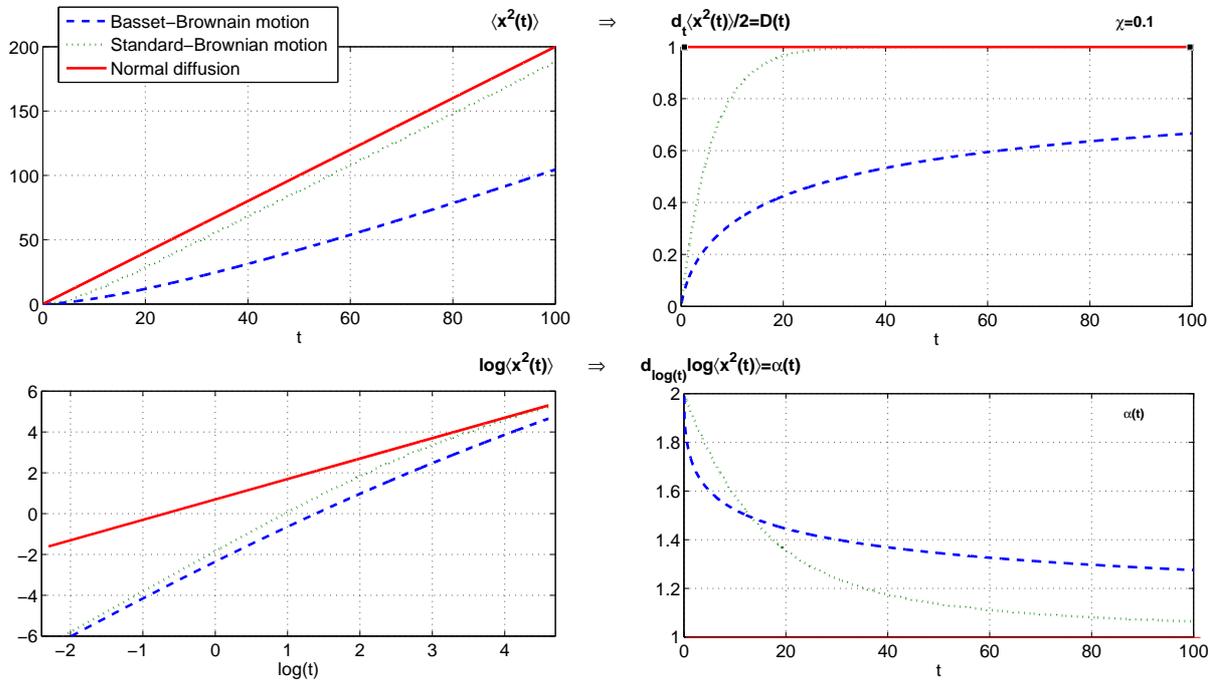}
\end{center}
\vskip -0.75truecm
 \caption{Plots   on the displacement variance for $\chi =0.1$}
 \label{figure 4}
 \end{figure}
 \vsp

\vsp
We will note that in all cases $D(t)$ is an increasing function of time, starting from $0$ at $t=0$
and  getting $1$ in the limit as $t\to \infty$.
On the other hand, $\alpha(t)$ is a decreasing function of time, starting from $2$ at $t=0$ (ballistic regime)
 and getting  $1$  in the limit as $t \to \infty$  (normal diffusion).   
%%%%%%%%%%%%%%%%%%%%%%%%%%%%%%
\vsp
%%To point out the regimes of fast diffusion
In our plots we scale  the time  $t$
 with the equilibrium relaxation $\sigma_e$ given by (3.1).
%% $$ \sigma_e  =  \sigma \, \l(1 +\frac{1}{2\chi}\r) \,,
%%  \q {\rm with} \q \chi=\frac{\rho_p}{\rho _f}\,. \eqno(3.1) $$
and, as usual,  we assume for $\chi$ lower  and lower values  
$\chi =  1, 0.5,  0.1$. 
%% Here we neglect the effect of added mass, that, as we have seen, is only limited
%% to change the scale of times. 
We easily recognize  that the retarding effect of the Basset force 
is larger for lighter particles, so for them the exponent $\alpha(t)$ remains over 1
for  longer times so that, in experiments of limited time duration,
this effect could be interpreted as fast diffusion.
%%%%%%%%%%%%%%%%%%%%%%%%
%

%%%%%%%
\section{Conclusions}
%%%%%%%%%
%%%%%%%%%%%%%%%%%%%%%%%%%%%%%%%%%%%%%%%%%%%%%%%%%%%%%%%%
The velocity autocorrelation
and the  displacement variance of a  Brownian particle
moving in an incompressible viscous fluid
 are known to be the fundamental   quantities
that characterize the evolution in time of the related stochastic process.
Here they
are  calculated
taking into account the  effects of {\it added mass} and both
{\it Stokes and Basset hydrodynamic forces} that
describe the friction  effects, %% in a viscous fluid,
respectively in the steady state and
in the transient state of the motion.
%% in the limit of vanishing Reynolds number.
%%%%%%%%
\vsp
 The explicit expressions of these quantities
versus time are computed %% where the  Mittag-Leffler functions
and compared with the respective ones for the classical Brownian motion.
\vsp
The effect of added mass is only  to modify the time scale,
that is the characteristic relaxation time induced by the Stokes force.
%%%
The effect of the Basset force, which is of hereditary type
namely  history-dependent,
is to perturb the white noise of the random force and
change the   decay character of the  {\it velocity
autocorrelation function} from pure exponential to power law
because of the presence of functions of {\it Mittag-Leffler}
type of order 1/2.
\vsp
Furthermore, the {\it displacement variance}
is shown to exhibit, for sufficiently long times,
the linear behaviour  which is typical of {\it normal diffusion},
with the same
diffusion coefficient   of the classical case.
However, due to the Mittag-Leffler functions, the  Basset history  force
induces  a very long retarding effect   in the
establishing of the linear behaviour,
 which could appear, at least for light particles, as a manifestation
 of anomalous diffusion of the fast type ({\it super-diffusion}).
%% To characterizes the diffusion regimes induced by the Basset
%% force  we show the behaviour of the instantaneous
%% diffusion coefficient.
%%%%%%
%%%%%%%
%%%%%%%%%%%%

%%%%%%%%%%

\section*{Acknowledgements}
%%%%%%%%%%
This research has been carried out in the framework
of the  programme
{\it Fractional Calculus Modelling},
see {\tt http://www.fracalmo.org}.
%% ESF Programme TAO:
%% "Transport in the Atmosphere and the Oceans".
F.M is grateful to the ISAC-CNR Institute
for  hospitality and to the late Professor 
Radu Balescu (Emeritus at the Universit\'e Libre de Bruxelles)
for inspiring discussions.
%%%%%%%%%%%%%%
%% \vfill\eject
%%%%%%%%%%
\section*{Appendix A}
%%%%%%%%%%%
Let us consider the {\it generalized Langevin equation} (3.10),
that we write as
$$   R(t) = m_e \, \left[ \dot V(t) + \, \gamma(t) *   V(t)\right]
\,,\eqno (A.1)$$
where $\cdot \null $ denotes time differentiation and $*$   time
convolution.
 The assumption of stationarity for the stochastic processes
along with the following hypothesis
$$ \langle \, R(t)\, \rangle= 0\,, \q
   \langle \, V(0) \, R(t) \, \rangle = 0\,, \q t>0\,,  \eqno(A.2)$$
allows us to derive, by using the Laplace transforms,
the two {\it fluctuation-dissipation theorems}

%% that, using the Laplace transforms, read
$$ \bar C_V(s) := \bar{{\langle\,  V(0) \, V(t)\, \rangle}}
   =  \frac{ \langle \, V^2(0)\, \rangle }{s + \bar \gamma (s)}
     \,, \eqno(A.3)$$
and
$$
 \bar C_R(s) := \bar{{\langle\,  R(0) \, R(t)\, \rangle}}
=  m_e^2 \,   { \langle \, V^2(0)\, \rangle}\,
     \bar \gamma (s)\,. \eqno(A.4)$$
Our derivation is alternative to
%% appears  simpler than
the original one by Kubo (1966) \cite{Kubo 66}
who used Fourier transforms; furthermore, it  appears useful
for the treatment of our {\it fractional Langevin equation}.
\vsp
Multiplying both sides of (A.1) by $V(0)$ and averaging, we
obtain
$${\langle\,  V(0) \, \dot V(t)\, \rangle}
  + \, \gamma (t) *   {\langle\,  V(0) \, V(t)\, \rangle} =0\,.
    \eqno(A.5) $$
The application of the Laplace transform to both sides of (A.5) yields
$$ s\, \bar{{\langle\,  V(0) \, V(t)\, \rangle}} \, - \,
    {\langle\,  V^2(0)\, \rangle} \, + \,\bar \gamma(s) \,
      \bar{{\langle\,  V(0) \, V(t)\, \rangle}} =0\,, \eqno(A.6)$$
from which  we just obtain (A.3).
\vsp
Multiplying both sides of (A.1) by $R(0)$ and averaging, we
obtain
$$ C_R(t) :=
       {\langle\,  R(0) \, R(t)\, \rangle}
            = m_e^2 \, \left[
    {\langle\, \dot  V(0) \, \dot V(t)\, \rangle}
  + \, \gamma (t) *  {\langle\, \dot V(0) \,  V(t)\, \rangle}\,\right]
\,. \eqno(A.7)$$
Multiplying both sides of (A.1) by $R(0)$ and averaging, we
obtain
$$ C_R(t) :=
       {\langle\,  R(0) \, R(t)\, \rangle}
            = m_e^2 \, \left[
    {\langle\, \dot  V(0) \, \dot V(t)\, \rangle}
  + \, \gamma (t) *  {\langle\, \dot V(0) \,  V(t)\, \rangle}\,\right]
\,. \eqno(A.7)$$
Noting that, by the stationary condition,
$$ {\langle\,  \dot V(0) \,  V(0)\, \rangle} =0 \,, \q
  {\langle\,  \dot V(0) \,  V(t)\, \rangle} =
 - \, {\langle\,  V(0) \,  \dot V(t)\, \rangle}\,, \eqno(A.8)$$
the application of  the Laplace transform to both  sides
of (A.7) yields
$$ \bar C_R(s) = m_e^2 \,
\left\{ s\, \bar {{\langle\, \dot V(0) \,  V(t)\, \rangle}}
        - \bar \gamma (s) \,
 \left[ s \, \bar {{\langle\,   V(0) \, V(t)\, \rangle}}-
     {\langle\,   V^2(0)\, \rangle}   \right]  \right\} \,. \eqno(A.9) $$
Since
 $$ \bar {{\langle\,  \dot V(0) \,  V(t)\, \rangle}}
 = - \, \bar {{\langle\,  V(0) \, \dot V(t)\, \rangle}}
 = -s \, \bar {{\langle\,  V(0) \,  V(t)\, \rangle}}
  + {\langle\,  V^2(0) \, \rangle}\,,  \eqno(A.10) $$
we get
$$ \bar C_R(s) = m_e^2 \,
\left\{ s\, \left[ -s\, \bar C_V(s)  +  {\langle\,  V^2(0) \, \rangle}
        - \bar \gamma (s) \, \bar C_V(s) \,\right]
 + \bar\gamma (s)\, {\langle\,   V^2(0)\, \rangle}   \right\} \,,
   \eqno(A.11) $$
from which, accounting for (A.3), we just obtain (A.4).
%%%%%%%%%%%%%% %%%%%%%%%%%%%
%% \vfill\eject
%%%%%%%%%%%%%%%%%%%%%%%%%%
\def\N{\bar N}  %%%%%%%%%%%%%
\def\ss{{s}^{1/2}} %%%%%%%
%%%%%%%%%%
\section*{Appendix B}
%%%%%%%%%%%%
In this Appendix we  report  the detailed manipulations
necessary to obtain the result (3.19) as Laplace inversion of (3.15).
For this purpose we need to consider the Laplace transform
     $$ \N(s) =
 {\frac{1}{s+ b \,\ss +1}}  \,, \q b = \sqrt{\beta}\,, \eqno(B.1)$$
and recognize that
$$ \frac{C_V (t)}{{\langle \, V^2(0)\, \rangle}}        = N(t/\sigma_e)
    \, \div \, \sigma _e\, \N( \sigma _e\,s) =
  \frac{1 }{s + \sqrt{\beta /\sigma_e } \, s^{1/2} +1/\sigma_e}
\,, \eqno(B.2)$$
where we have used the sign $\div$ for the juxtaposition of
a function depending on $t$ with its Laplace transform depending on $s\,.$
The required result is  obtained by expanding
 $\N(s)$ into  partial fractions and then inverting.
Considering the two roots $\lambda _\pm$ of   the polynomial
$P(z) \equiv z^2 + b \,z  +1\,$ with $z=s^{1/2}\,,$
 we must treat separately the following two cases:
 $ i)\;\; 0<b<2\,, \q{\rm or}\q  2<b<3\,,$  and
  $ii)\;\; b=2\,,$
 which correspond to two distinct roots ($\lambda _+ \ne \lambda _-$),
 or two coincident roots ($\lambda _+ \equiv \lambda _-= -1$),
 respectively.
%% \vsp
We obtain
\vsh
\vsn
   $ i)\q   b \ne 2 \, \Longleftrightarrow\,
  \beta \ne 4 \,, \; \chi  \ne 5/8\,,$
$$  \N(s) = \frac{1}{s+ b \,\ss +1}
 = \frac{A_+}{\ss\,(\ss -\lambda_+ )} -
    \frac{A_-}{\ss\,(\ss -\lambda_- )} \,,
\eqno (B.3)  $$
with
$$\lambda _\pm = \frac{-b \pm (b^2 -4)^{1/2}}{2}=\frac{1}{\lambda _\mp}
     \,, \quad
   A_{\pm} =  \frac{ \lambda _\pm}{\lambda _+  - \lambda _- }\,;
 \eqno(B.4)
$$
\vsn   $ii) \q b= 2 \, \Longleftrightarrow\,
   \beta = 4 \,, \; \chi  = 5/8\,,$
$$  \N(s) = \frac{1}{s+ 2 \,\ss +1}  =\frac{1}{(\ss +1)^2}
 \,.\eqno (B.5)$$
The Laplace inversion of  (B.3) and (B.5)
turns out, see below,
$$N(t) = \left\{
\begin{array}{ll}
i)& A_+\,E_{1/2}(\lambda_+\sqrt{t})-A_-E_{1/2}(\lambda_-\sqrt{t}),\\
ii) &  (1+2t)\, E_{1/2}\,(-\sqrt{t})  - 2\, \sqrt{t/\pi}
\end{array}
\right.
   \eqno (B.6)
   $$
where
 $$ E_{1/2} (\lambda \,\sqrt{t})  =
   \sum_{n=0}^\infty  \frac{\lambda ^n\, t^{n/2}}{\Gamma (n/2+1)}
     =   {\rm e}^{\ds \lambda ^2\,t}\, {\rm erfc} (-\lambda \,\sqrt{t})
\eqno(B.7)
$$ denotes the {\it Mittag-Leffler  function} of order $1/2\, $
and erfc denotes the complementary error function.
In view of (B.1-2),  equation (B.6) is equivalent to (3.19).
\vsp
Let us first recall  the essentials of the generic Mittag-Leffler
function in the framework of the  Laplace transforms.
The Mittag-Leffler function $E_\alpha (z)$ with $\alpha >0\,, $
so named from the great Swedish mathematician
who introduced it at the beginning of
this century,
is defined by the following series representation,
valid in the whole complex plane,
$$
E_\alpha (z) =  \sum_{n=0}^\infty \frac{z^n}{\Gamma (\alpha n+1)}
\,,\q \alpha > 0\,, \q z\in \CC\,. \eqno(B.8)$$
It turns out that
$E_\alpha (z) $ is an {\it entire function},
 of order $\rho =1/\alpha \,$ and type $1\,, $
which provides a  generalization of the
  exponential function.
%% because of the substitution of
%% $n! = \Gamma (n+1)$ with $(\alpha n)!=\Gamma (\alpha n+1)\,.$
%% \vsp
%% Particular cases of (B.1), from which elementary functions
%% are   recovered,  %%% easily recognized
%% are
%% $$ E_2\l(+z^2\r) =  \cosh \, z\,, \qq
%%   E_2\l(-z^2\r) =  \cos \, z\,, \qq z\in \CC \,,   \eqno(B.3)$$
%% and
%% $$ E_{1/2} (\pm  z^{1/2}) =
%%     \e^{\ds z} \, \l[ 1+{\rm erf}\, (\pm z^{1/2})\r ] =
%%   \e^{\ds z} \, {\rm erfc} \, (\mp z^{1/2})\,,\q z\in \CC \,,
%%  \eqno(B.4)$$
%% where erf (erfc) denotes the (complementary) error function
%% defined as
%% $$  {\rm erf} \,(z) =  \frac{2}{ \sqrt{\pi}}\,\int_0^z \e^{\ds -u^2}\,du
%% \,,  \q   {\rm erfc} \,(z) =   1 - {\rm erf}\, (z)\,,
%%  \q z\in \CC\,. $$
%% In (B.4) for $z^{1/2}$ we mean the principal value of the
%% square root of $z$  in the complex plane cut along the the
%% negative real axis. With this choice $\pm z^{1/2}$
%% turns out to be positive/negative  for $z \in \RR^+$.
%%%%%%
\vsp
%%%%%%%%%%%
     The Mittag-Leffler function  is connected to the
Laplace integral through the equation
$$
\int_0^\infty \!\!\! \e^{-u} \, E_\alpha \left(u^\alpha \,z\right) \, du
   = \frac{1}{1-z}\,,
  \q \alpha >  0
   \,.\eqno(B.9)$$
%% The integral at the L.H.S. was evaluated by Mittag-Leffler who showed
%% that the region of its convergence contains the
%% unit circle and is bounded by the line
%% ${\rm Re}\, z^{1/\alpha } =1$.
This integral is fundamental in the  evaluation of
the {Laplace transform} of %% the functions
$E_\alpha \left(\lambda \, t^\alpha \right)$
with $\lambda  \in \CC\, $  and $t \ge 0\,. $
%% \vsp
Putting  in (B.9)
$u=st$ and $u^\alpha \, z = \lambda \, t^\alpha\,, $
we get the following Laplace transform pair
$$
 E_\alpha \left(\lambda \, t^\alpha \right)
   \div \frac{s^{\alpha -1}}{s^\alpha -\lambda} \,,
 \q Re \, s > |\lambda |^{1/\alpha }
\,. \eqno(B.10)$$
We note that, up to our knowledge, in the handbooks containing tables for
the Laplace transforms, the Mittag-Leffler
function is ignored so that the transform pair (B.10)
does not appear
if not in the special case $\alpha =1/2\,.$
%% written, however, in terms of the exponential
%% and complementary error functions.
%% see \eg [71].
In fact, in this case
we  recover from (B.10)
the basic Laplace transform pair
%%%%%%%%%%%%%%%%%%%%%
\def\ss{{s}^{1/2}}   %% for LAPLACE TRANSFORMS
\def\st{{\sqrt{t}}}
\def\lst{{\lambda \,\st}}
\def\Et{{E_{1/2}(\lst)}}
%%%%%%%%%%%%%%%%%%
$$
\frac{1}{\ss\,(\ss -\lambda)} \,\div \, \Et
   \,,  \eqno(B.11) $$
where the Mittag-Leffler function can be expressed in terms of
known functions, as shown in (B.7).
As an exercise
we can derive from (B.11)
the following
transform pairs and consequently  the result (B.6):
$$
 \frac{1}{\ss -\lambda}= \frac{1}{\ss} +\frac{\lambda}{\ss\,(\ss -\lambda)}
 \,\div \,       \frac{1}{\sqrt{\pi\,t}}
 + \lambda \, \Et
    \,,  \eqno(B.12)$$
$$       \frac{1}{\ss\, (\ss -\lambda )^2}  =
   -2\, \frac{d}{ds}\,\left(\frac{1}{\ss -\lambda }        \right)
 \,\div \,  2\,\sqrt{\frac{t}{\pi}}
 + 2\,\lambda \, t\, \Et\,,  \eqno(B.13) $$
$$
\begin{array}{llll}
\displaystyle
 \frac{1}{(\ss -\lambda)^2}& \displaystyle = 
 & \displaystyle \frac{1}{\ss\, (\ss -\lambda)}+\frac{\lambda}{\ss\, (\ss -\lambda )^2 } \\
&\div& 2\lambda\sqrt{\frac{t}{\pi}}+(1+ 2\,\lambda^2\,t)\, \Et
\end{array}
\eqno(B.14) $$
%%%%%%%%%%%% THE END OF APPENDIX
% BIBLIOGRA{PHY
% \input{mainardi-mura-tampieri_biblio.tex}
%%%%%%%%%%%%%%%%%
%%%%%%% mainardi-mura-tampieri_biblio.tex  18.12.09

\end{document}